\documentclass[aps,prl,preprint,showkeys]{revtex4}

\begin{document}

\title{Measuring Time of Flight Using Satellite-Based Clocks}

\author{Ronald A.J. van Elburg}
\email{RonaldAJ(at)vanelburg.eu}
\affiliation{Department of Artificial Intelligence, 
Faculty of Mathematics and Natural Sciences,  
University of Groningen, 
P.O. Box 72, 
9700 AB, Groningen, 
The Netherlands}
\date{31 October 2011}

\begin{abstract}
Considering the OPERA neutrino-velocity measurement from the point of view of a GPS satellite we find that the detector at Gran Sasso has a velocity component in the order of $10^{-5}c$ towards the neutrino emission location at CERN. On GPS-receivers this translates into first-order Doppler terms, therefore a correction is required for, among other things, this ephemeris-and-location-dependent relativistic effect. To ensure correct time-of-flight measurements using satellite-based clocks we propose to extend their calibration procedures with an explicit check on these relativistic corrections.
\end{abstract}

\keywords{clock calibration, time measurement, OPERA experiment, neutrino velocity}

\maketitle
This paper is a response to the extraordinary finding of superluminal neutrino velocities presented in the OPERA experiment~\cite{OPERA}. The OPERA collaboration presented this result in order ``to invite scrutiny from the broader particle physics community'' and to make ``sure that there are no other, more mundane, explanations''~\cite{OPERAPRESS2011}. The ICARUS collaboration~\cite{ICARUS2011} tested predictions by Cohen and Sheldon~\cite{Cohen2011} on the effect of superluminal velocities of neutrinos on the neutrino energy distributions and found results which are inconsistent with superluminal neutrino velocities. In view of this conflicting result, and the many confirmations of special and general relativity in the past it is important to rule out  potential technical explanations for the outcome of the OPERA time-of-flight experiment. When judging such potential explanations the question should not be whether they are simple or complicated, but rather whether we find evidence that these explanations have been convincingly checked and hence ruled out. 

An analysis of the OPERA experiment should include an analysis of potential pitfalls in all the technologies employed. Due to the central importance of the clocks this certainly includes a scrutiny of GPS-based time measurement. In this light it is important to know that a recent review of common-view GPS~\cite{Levine2008} cautions that ``if the transmitter is moving'' extra ``contributions to the error budget'' arise. How to take into account the different ``contributions to the error budget'' is elucidated by another paper. This paper explains that the clocks in the GPS satellites are adjusted for the second-order Doppler effect and a blueshift due to difference in gravitational potential between a satellite and an observer on earth. However, ephemeris-and-location-dependent corrections, that is corrections varying with satellite position and receiver location, such as the first-order Doppler effect and a frequency offset due to eccentricity, are to be applied by the receiver~\cite{Levine1999}. 

In the description of the OPERA experiment~\cite{OPERA} the present author found no account of a check on receiver-side relativistic GPS corrections. Likewise, the author found no discussions of these first-order effects in the accompanying calibration reports~\cite{PTB2011,METAS2008}. In fact, the calibration report~\cite{PTB2011} shows that the timing signal is obtained from a GPS-driven rubidium clock and not from the calibrated GPS receiver. Corrections to the GPS-driven rubidium clocks are computed every second to maintain synchrony between CERN and Gran Sasso, but these clocks still run at their original frequencies between two corrections.  Consequently, if these GPS-based clocks use the time-dilation-corrected satellite clock without the first-order corrections, then the whole experiment is actually set in the satellite reference frame. 

Although the speed of light is invariant under such a change of reference frame, special relativity does not preserve distance and time separately. In fact, as first illustrated by the interference pattern in the Michelson-Morley experiment~\cite{Michelson1887}, to render experimental outcomes reference-frame independent, space and time need to be transformed according to joint Lorentz transformations rendering the speed of light invariant. The Lorentz transformations show that for the times of flight between the different parts of an experimental set-up it is not sufficient to apply the Lorentz contraction, it is also necessary to take into account the movement of these parts with respect to the reference frame used. 

This applies in the same way if we want to determine a particle's time of flight in an earth-based experiment like the OPERA experiment~\cite{OPERA} in which we use a clock attached to a moving reference frame such as that of a non-stationary satellite.  Let us use the well known case of photon moving freely in vacuum and hence moving at the speed of light, to estimate the time discrepancy for a particle with a velocity close to the speed of light sent from CERN to Gran Sasso. Given the short duration of the experiment the satellite reference frame is well approximated by an inertial reference frame. The source and detector are separated by a fixed distance $S_{b}=|{\bf x}_{Gran Sasso}-{\bf x}_{CERN}|$ in their baseline reference frame. 
The time of flight $\tau_{b}$ for a photon in the baseline reference frame is now simply given by,
\begin{equation}
\tau_{b}=\frac{S_{b}}{c}.
\end{equation}
For further simplification we will assume that the velocity vector ${\bf v}$ of the satellite is strictly parallel to the baseline, i.e. ${\bf v}=v {\bf b}$ where ${\bf b}$ is a unit vector pointing from CERN towards Gran Sasso: ${\bf b}=({\bf x}_{Gran Sasso}-{\bf x}_{CERN})/S_{b}$. We shall later show that $v$ is in the order of $10^{-5}c$. After a photon is emitted, two movements in the satellite reference frame are observed: the photon travels towards the detector at the speed of light and the detector at Gran Sasso moves towards the photon-emission location originally at CERN with a velocity $v=|-{\bf v}|$.  The detector movement is towards the emission location originally at CERN because the satellite moves from west to east along its orbit with a strong velocity component parallel to the path of the neutrinos. Consequently, the time required for the photon to reach the detector is shorter than the (Lorentz-contracted) distance separating the source and detector. This is entirely due to the movement of the detector at Gran Sasso towards the emission location originally at CERN in the satellite's reference frame. This is true despite the fact that in the baseline reference frame the distance separating the source and detector is exactly equal to the distance separating the emission event at CERN and the detection event at Gran Sasso. 

We now calculate the photon time of flight in the satellite reference frame, and compare it to the time-of-flight estimate for photons in the baseline reference frame. If we identify $\bf{b}$ with the $\bf{x}$ direction, then in the baseline reference system we can define the coordinates of the emission event at CERN to be  $(x_b,t_b)_e=(0,0)$ and the coordinates of the detection at Gran Sasso to be $(x_b,t_b)_d=(S_b,\tau_b)$. Using the Lorentz transformation~\cite{BPC1} gives us $(x_s,t_s)_e=(0,0)$ for the coordinates of the emission event and 
\begin{eqnarray}
 (x_s,t_s)_d &=& \gamma \left(S_b-v\tau_b,\tau_b-\frac{v}{c^2} S_b\right) \nonumber\\
 &=&\frac{S_b}{\gamma(c+v)} (c,1)
\end{eqnarray}
for the coordinates of the detection event in the satellite reference frame, where $\gamma={1}/{\sqrt{1-{\bf v}^2/c^2}}$. From the expression above we can simply read off the time of flight in the satellite reference system:
\begin{equation}
\tau_{s}  = \frac{S_{b}}{\gamma} \frac{1}{c+v}= \frac{S_{s}}{c+v}.
\label{eq:TOFs}
\end{equation}
We should stress that here $S_s$ is the Lorentz-contracted baseline distance $S_b$, and not the coordinate of the detection event. From this result we further recover that under the Lorentz transformation the ratio between the time of flight and the distance travelled indeed equals the speed of light in both reference frames, i.e $S_b/\tau_b=c/1=c$.
 
As mentioned above, ${S_{b}}/{\gamma}$ corresponds to the Lorentz-contracted distance $S_{s}$ between the source and detector in the satellite reference frame. In the satellite reference frame we can easily interpret the factor ${1}/{(c+v)}$. In this reference frame the detector moves with a velocity $-{\bf v}$, i.e. it moves towards the emission location originally at CERN. To find the distance traveled by the photon, we can subtract the distance traveled by the detector during the time of flight from the total distance separating the source at CERN and the detector at Gran Sasso. In the satellite reference frame the photon will therefore have to cover a shorter distance $S_{s} - v \tau_{s}$ leading to a shorter time of flight $\tau_{s}$ in the satellite reference frame. From this we obtain $c \tau_{s}=S_{s} - v \tau_{s}$, which also leads us to the correct answer as given in equation~\ref{eq:TOFs}.

Let us now consider a potential pitfall in the use of satellite-based clocks in the OPERA experiment. A GPS satellite's clock is configured to provide GPS receivers with a signal that on average provides the correct clock speed, i.e. a signal that is corrected for time dilation. However, the necessary ephemeris-and-location-dependent relativistic corrections are the responsibility of the GPS receiver. The authors of the OPERA paper~\cite{OPERA} and accompanying calibration reports~\cite{PTB2011,METAS2008} do not explicitly mention these corrections. Hence we cannot rule out the possibility that the OPERA experiment is not measuring time of flight in the baseline reference system, but instead the satellite-reference-system time of flight only corrected for the Lorentz dilation and not for the ephemeris-and-location-dependent relativistic corrections:
\begin{equation}
\tau_{o} =\gamma \tau_{s} =\frac{S_{b}}{c+v}.
\end{equation}
For a photon the difference between the baseline time of flight $\tau_{b}$ and the observed time of flight $\tau_{o}$ would then be given by
\begin{equation}
\epsilon=\tau_{b}-\tau_{o}=\frac{S_{b}}{c}\left(1-\frac{c}{c+v}\right)=\frac{S_b}{c+v} \frac{v}{c}\approx \tau_{b} \frac{v}{c}.
\label{eq:epsilon}
\end{equation} 
Note that the corrections related to detector movement are first-order in $v/c$.  Corrections related to time dilation or Lorentz contraction, i.e. involving $\gamma$, only contribute corrections of the order ${\bf v}^2/c^2$, and from our upcoming estimate for $v\approx 10^{-5} c$ we know these contributions to be in the order of $10^{-10}$.

We now calculate the quantity $\epsilon$ for a photon and show that it is in the order of the observed deviation of the neutrino time of flight from the expected time of flight. The clocks in the OPERA experiment are orbiting the Earth in GPS satellites. The orbits of these satellites are at an altitude of $20.2 \cdot 10^6 {\rm \ m}$ from the Earth's surface in a fixed plane inclined $55^\circ$ from the equator with an orbital period of 11 h 58 min~\cite{Lombardi_2001}. This implies that they fly predominantly west to east when they are in common view of both the source location (CERN) and the detector location (Gran Sasso), i.e. they have a velocity component in the same direction as the neutrino's velocity. In fact when in common view, part of their orbit is parallel to the CERN-Gran Sasso line. The radius of a GPS satellite's orbit is obtained by adding its altitude to the radius of the Earth,  $6.4 \cdot  10^6 {\rm \ m}$, which in this case yields a total radius of $26.6\cdot10^6 {\rm \ m}$. The velocity of the GPS satellites is therefore approximately 
\begin{equation}
v=2\pi R/T = 2\pi\cdot 26.6\cdot10^6 {\rm \ m}/(11.58 \cdot 60 \cdot 60 {\rm \ s})= 4.0\cdot10^3 {\rm \ m/s}\approx 1.3 \cdot 10^{-5}c,
\end{equation}  
where $c$ is the speed of light $c\approx 3.0\cdot 10^8 {\rm \ m/s}$.
Returning to equation~(\ref{eq:epsilon}) and using $S_{b}=7.3\cdot10^5{\rm \ m}$ , we obtain:
\begin{equation}
\epsilon = 32 {\rm \ ns}.
\end{equation}   
Thus, when two GPS receivers are used which do not implement ephemeris-and-location-dependent relativistic corrections, the observed time of flight would be $32$ ns shorter than the actual time of flight in the worst case scenario described here. In view of the reported statistical $\pm 6.9$ ns and systematic $\pm 7.4$ ns uncertainties, this is a large part of the reported 60 ns in the OPERA paper.

To rule out ephemeris-and-location-dependent relativistic effects as a potential source of error, the OPERA experiment should be examined to identify potential locations where such effects need correction. As most of the corrections that the OPERA team documented are estimated using local baseline-based clocks, these corrections do not need extra adjustment. However, the GPS receivers which provide the time stamps and subsequent data-processing steps need to be critically examined. Synchronization, whether or not using GPS Common View, does not necessarily ensure that the clocks involved run at the correct frequency. 

\section*{Conclusion}
In the OPERA paper~\cite{OPERA} and accompanying calibration reports~\cite{PTB2011,METAS2008} we found no evidence that the GPS-based clocks are convincingly checked for first-order Doppler effects. We showed that taking these first-order Doppler effects into account gives a correction to the OPERA experiment which can explain a large part of the discrepancy between the time of flight the OPERA team observed and the time of flight expected. 

To simplify our presentation we have presented our analysis in the reference frame of a GPS satellite. An alternative presentation in the CERN-Gran Sasso reference frame will yield the same corrections. Furthermore, the calculation presented above also contains some simplifying assumptions. A full treatment, which by necessity relies on the experimental data, should take into account, for instance, the varying angle between the GPS satellite's velocity vector and the CERN-Gran Sasso baseline. We expect that such a full treatment will yield a somewhat smaller value for the average correction. In addition, such a full analysis should be able to predict the correlation between the GPS satellite position(s) and the observed time of flight, and should take into account all technical details of the common-view GPS-based timing methodology used. Our treatment holds for the component of the satellite's velocity parallel to the CERN-Gran Sasso baseline. The transverse component only gives ${\bf v}^2/c^2$ effects which can be neglected for our current purpose.

We know from the theory of special relativity~\cite{Einstein1905} that time is reference-frame specific. Accordingly, this paper stresses that we have to take into account how different clocks are moving. Some of the complications introduced can be avoided by using geostationary clocks. It is important, however, to realize that the corrections described in this paper are specific to the experiment, as they vary with the orientation of the baseline with respect to the satellite's path. There is no {\it a priori} reason to expect that synchronization between clocks can account for effects related to the orientation of the experiment. In addition, the GPS system was originally designed for spatial localization~\cite{Fliegel1996}. This application essentially depends on relative rather than absolute timing and is less sensitive~\cite{Kalman1960} to the effect of special relativity than its usage in the OPERA experiment, i.e., the absolute synchronization of two clocks. 

It is evident that the issue of corrrect GPS timing needs to be satisfactorily resolved, not only for the correct interpretation of the OPERA results, but also for the implementation of clock synchronization through common-view GPS in similar large baseline experiments. In our opinion the clock calibration process should include a step which explicitly verifies that the necessary corrections are applied in the GPS receiver and in the accompanying post-processing steps. 

\begin{acknowledgments}
The author is a member of the Sensor City (Assen,The Netherlands) project team which is funded by the European Union, European Regional Development Fund, the Dutch Ministry for Economical Affairs and the Northern Netherlands Provinces. The author wishes to thank Tjeerd Andringa for discussions during the early phase of this work and Lambert Schomaker and Theo Nieuwenhuizen for support and suggestions in the concluding phase. 
\end{acknowledgments}


\end{document}